# Physical properties of thermoelectric zinc antimonide using first-principles calculations

Philippe Jund[a], Romain Viennois[a], Xiaoma Tao[b], Kinga Niedziolka[a] and Jean-Claude Tédenac[a]

[a] Institut Charles Gerhardt, Université Montpellier 2, Pl. E. Bataillon CC15003
34095 Montpellier, France

[b] College of Physical Science and Technology, Guangxi University
Nanning, 530004, P. R. China

Abstract:

We report first principles calculations of the structural, electronic, elastic and vibrational properties of the semiconducting orthorhombic ZnSb compound. We study also the intrinsic point defects in order to eventually improve the thermoelectric properties of this already very promising thermoelectric material. Concerning the electronic properties, in addition to the band structure, we show that the Zn (Sb) crystallographically equivalent atoms are not exactly equivalent from the electronic point of view. Lattice dynamics, elastic and thermodynamic properties are found to be in good agreement with the experiments and they confirm the non equivalency of the zinc and antimony atoms from the vibrational point of view. The calculated elastic properties show a relatively weak anisotropy and the hardest direction is the y direction. We observe the presence of low energy modes involving both Zn and Sb atoms at about 5-6 meV, similarly to what has been found in $Zn_4Sb_3$ and we suggest that the interactions of these modes with acoustic phonons could explain the relatively low thermal conductivity of ZnSb. Zinc vacancies are the most stable defects and this explains the intrinsic p-type conductivity of ZnSb.

1. Introduction

Among the sustainable energy sources, thermoelectricity has attracted the attention during the last years because of the need to solve the urgent energetic problems and of recent advances in the search of more efficient thermoelectric materials. Indeed, since the middle of the nineties, new thermoelectric materials with large Figures of Merit Z have appeared and thus higher efficiencies could be obtained [1]. However, for high-temperature applications on a large scale, one should take into account not only the thermoelectric efficiency but also some other important factors such as:

- the abundance, cost, toxicity of the elements used in the thermoelectric material,

- the thermodynamic and mechanical stability of the material in the desired temperature,

- the possibility to have both n and p type conductors which is connected to the doping and to the role of the intrinsic point defects.

Therefore, in the quest of new thermoelectric materials, it is necessary not only to study the properties directly related to their thermoelectric characteristics (electronic properties and lattice dynamics) but also to study their elastic properties, their thermodynamic properties and their stability (both pure and doped phases). This is the scope of the present paper for one of the most promising thermoelectric materials already known for quite some time: ZnSb.

ZnSb is one of the best thermoelectric compounds in the important temperature range between 400 and 600 K for which only a limited number of efficient thermoelectric materials is known. With zT = 1.4 at 600 K, the most efficient material known to us is $Zn_4Sb_3$ [1,2], however it is metastable [3-5] and can only be of p-type [1,6]. Other possibilities concern tellurides such as LAST and TAGS [1]; however, their use can only be very limited due to the weak abundance and the toxicity of tellurium. Therefore, after excluding these compounds, the most efficient material is orthorhombic ZnSb in this temperature range [1]. This compound is a slightly anisotropic semiconductor with a bandgap of about 0.5 eV [7-12]. Several experimental studies were dedicated to this material during the first period of intense activity on thermoelectric materials [7-12]. During this time, the best zT was around 0.6 for the p-type doped material [1,7,8,10]. Some studies of n-doped compounds with impurities such as In, Te or Se were also reported [9,11] but the zT was too low. In parallel there were also several studies of the thermodynamic properties of ZnSb and CdSb-ZnSb alloys with the same structure [12-17]. Since the nineties, due to the discovery of very good thermoelectric properties in $Zn_4Sb_3$ [2] the main studies of Zn-Sb systems were dedicated to this last compound and the study of orthorhombic ZnSb became confidential. The main reason for which the thermoelectric properties of $Zn_4Sb_3$ are better than in ZnSb is its two times smaller thermal conductivity [1]. However, after more than one decade of effort it is still not possible to improve the stability of $Zn_4Sb_3$ and no n-doped material has been found so far [1,6].

It is therefore time to thoroughly study ZnSb in order to improve its thermoelectric properties. Recently, it has been reported that it is possible to reduce successfully the thermal conductivity of ZnSb by nanostructuring [18,19] and increase its zT up to about 1 [19]. It has also been reported that the n-doped compound with tellurium can have similar electronic properties (and hence zT) to the p-doped material [20] (recent experimental studies have explored other n-type doping but without significant success [21]). These recent results open new perspectives for the orthorhombic ZnSb compound without the stability and doping problems of $Zn_4Sb_3$. In the past only a few papers have dealt with the numerical study of the physical properties of ZnSb [22,23] but during the last two years several studies have been devoted to this topic [24-27]. In view of possible application in high temperature thermoelectric generation, it is thus necessary to deepen our understanding of the overall physical properties of ZnSb and not only of its physical properties directly related to its thermoelectric properties. The aim of the present paper is to propose a full ab initio study of the electronic properties (in addition to the study of Benson at al. [27]), the lattice dynamics, the thermodynamic and elastic properties and the stability of native point defects in ZnSb in order to better probe its potential for future thermoelectric applications.

2. Computational details

First-principles calculations are performed using the projector augmented-wave (PAW) method [28-29] within the local density approximation (LDA) and the generalized gradient approximation (GGA), as implemented in the highly-efficient Vienna Ab initio Simulation Package (VASP) [30]. The calculations employed the Perdew-Bucke-Ernzerhof (PBE) exchange-correlation functional within the GGA [31]. We have used a plane-wave energy cutoff of 500 eV held constant for all the calculations. For the relaxation and electronic structure calculations in the primitive cell, Brillouin zone integrations are performed using Monkhorst-Pack k-point meshes [32], with a k-point sampling of 15x15x15 and the tetrahedron method with Blöchl correction [33] is used in the present calculations. The total energy is converged numerically to less than $1\times10^{-6}$ eV/unit. After structural optimization, calculated forces are converged to less than $10^{-3}$ eV/Å. For the calculation of the elastic constants of ZnSb, the procedure is similar to the one described in detail in Ref.

[37] and is therefore not recalled here.

We use the Vinet equation of state to obtain the equilibrium volume ($\Omega_0$), and the total energy (E) [34]. Lattice dynamics calculations were done using the frozen phonons method in the supercell approach as discussed by Parlinski [35]. We calculate the Hellmann-Feynman forces in a 2x2x2 supercell of 128 atoms with a precision better than $10^{-5}$ eV/Å after a first step of ionic relaxation in the supercell and then the dynamical matrix is diagonalized using Parlinski's Phonon code [35]. From these phonon calculations, the thermodynamic properties and the Atomic Displacement Parameter (ADP) tensors of each atomic type are calculated (see ref. [35] for more details.

For the defect calculations, we also used a 2x2x2 supercell with the accuracies on the energy and the forces mentioned previously for the primitive cell but with a k-point sampling of 3x3x3 (similar to the calculation of the Hellmann-Feynman forces).

## 3. Results and discussion

### 3.1 Crystal structure and formation energy

ZnSb adopts an orthorhombic structure (Space Group Pbca, n°61) under ambient conditions with Zn and Sb atoms occupying the 8c Wyckoff positions leading to a unit cell containing 16 atoms. As stressed by other authors the structure can be described as an arrangement of interconnected planar rhomboid rings $Zn_2Sb_2$ as shown in Fig.1. The calculated lattice constants (a, b, c) and formation enthalpies have been listed in Table 1 together with the available experimental data [38].

Globally the calculated lattice constants are overestimated by at most 1.5% which is certainly due to the use of the GGA since it is well known that this approximation overestimates the lattice constants or the equilibrium volume [39]. The contrary is true within the LDA which does not improve the calculated lattice constants (Table 1). Nevertheless in both cases, the calculated c/a and b/a ratio (1.31 and 1.246 for GGA and 1.307 and 1.241 for LDA) are very close to the experimental ones (1.305 and 1.248).

The formation enthalpy of ZnSb in eV/atom can be calculated with the following equation:

$$\Delta H(ZnSb) = E(ZnSb) - (N_{Zn}E(Zn)/N_{tot} + N_{Sb}E(Sb)/N_{tot}) \qquad (1)$$

where $E(ZnSb)$, $E(Zn)$ and $E(Sb)$ are the equilibrium first-principles calculated total energies (in eV/atom) of the corresponding ZnSb compound, of Zn with hcp ($P6_3/mmc$) structure and of Sb with hexagonal structure (R-3m), respectively. $N_{Zn}$ is the number of zinc atoms and $N_{Sb}$ the number of Sb atoms.

Concerning the formation enthalpy of ZnSb the experimental values (from -0.07 to -0.09 eV/atom) [12,16-17] are relatively small and the DFT-GGA calculations are overestimating these results (about -0.035 eV/atom). We have performed an LDA calculation, but the agreement with experiment is even worse (-0.002 eV/atom) so clearly LDA calculations don't improve the structural results.

3.2 Electronic properties

We have done the full analysis of the electronic properties (including the Bader charge analysis) but since our results are similar to previously published results ([23], [26], [27]) we present here only the band structure (Fig. 2) and the partial density of states (DOS) of the different atoms (Fig. 3) since they either provide new informations or are necessary for the discussion of the elastic and thermodynamic properties.

The calculated band structure shows an indirect band gap of about 0.05 eV (similarly to what was found by Benson et al. [27] with the PBE functional) which is notably smaller than the experimental gap of 0.53eV [7]. This is a known flaw of the DFT-GGA description of semiconductors which systematically underestimates the band gap. To improve this, other more sophisticated exchange-correlation functionals should be used as shown in [26]. However, the main features of the band structure remain the same and can be related to the anisotropy of the physical properties, as will be discussed later.

In order to correlate the structure build on $Zn_2Sb_2$ rings and the DOS, we have analyzed the total DOS of the 8 Zn atoms and the 8 Sb atoms in the orthorhombic unit cell. A priori they should be equivalent for each atomic type since they occupy the same Wyckoff positions but this is not the case as shown in Fig.3. We find four different zinc atoms and four different antimony atoms corresponding to the four different rings embedded in the unit cell (see Fig. 1). In each ring the DOS of the 2 zinc atoms and the DOS of the 2 antimony atoms are similar but they are (slightly) different from one ring to another. The differences are small but they can *a priori* not be attributed to calculation errors since the DOS of the equivalent atoms are STRICTLY equal.

We have made several checks with different minimization methods, with different accuracies and with the experimental (non relaxed and fully symmetric) atomic positions and we find always the same result: the atoms go by pairs that have different electronic properties. This shows that from the electronic point of view the four rings to which the 16 atoms in the unit cell belong are not strictly equivalent. We will see below that these differences are also detectable in the vibrational properties.

3.3 Elastic properties

The knowledge of the elastic constants (both experimentally and theoretically) of a thermoelectric material is very important since the thermo-mechanical constraints can be very important during the lifetime of this material and thus it can be useful to predict its aging behaviour.

We have determined the bulk modulus from the fit of the E=f(V) data using the Vinet equation, as we have done for other materials previously [37]. With this method, we find a bulk modulus $B_{EOS}$ of 47.18 GPa and a pressure derivative $B_{EOS}$' of 5.35. We have also determined all the elastic constants of ZnSb using the method described in detail in [37]. The strains used for the calculations together with the induced stresses are shown in Table 2. The obtained elastic constants are shown in Table 3 together with experimental results from [14] and [40]. The mechanical stability of an orthorhombic system implies that [41]: $C_{11} > 0$, $C_{22} > 0$, $C_{33} > 0$, $C_{44} > 0$, $C_{55} > 0$, $C_{66} > 0$, $(C_{11} + C_{22} - 2C_{12}) > 0$, $(C_{22} + C_{33} - 2C_{13}) > 0$ and $[C_{11}+C_{22}+C_{33}+2(C_{12} + C_{13} + C_{23})] > 0$. As can be seen in Table 3, all these conditions are indeed fulfilled in our case.

The bulk modulus $B_H$ and the shear modulus $G_H$ are the arithmetic averages for powder samples from the Reuss and Voigt values obtained following the Hill's procedure as described in our previous work in the case of other symmetries [37] and by Ravindran et al for the orthorhombic case [41]. In the literature, we have found two sets of experimental values [14,40]. Our results agree best with the ones of [40]. This is not surprising since the resonant ultrasonic spectroscopy measurements in [40] are more accurate (error estimated to 0.5 % for the diagonal ($C_{ii}$) tensor components and 3 % for the non-diagonal ($C_{ij}$) tensor components) than the determinations made in [14] (error estimated to 2 % for the longitudinal velocity and 10 % for the transverse velocity). In addition, the authors in [14] note that the ultrasound is markedly damped in the sample they have investigated. Therefore, in the following, we will consider only the experimental results of [40]. We note that these experimental results agree qualitatively well with our computed results notably for the different directions, but we observe a systematic underestimation of the calculated elastic constants $B_H$, $G_H$ and $E_H$ of about 15 %. This is mostly related to the larger calculated lattice parameters compared to the experimental ones and the fact that we find the same c/a and b/a ratio than experimentally. This is why our calculations reproduce well the experimental tendency. We note that $B_H$, the powder-averaged bulk modulus of about 54.75 GPa found from the experimental data of Balaziuk et al [40] is larger than the value of 50.5 GPa found for $Zn_4Sb_3$ [42]. This shows that the naïve use of a weighted sum of the bulk modulus of zinc and antimony as proposed by Triches et al. [42], which would give a bulk modulus of only 49.05 GPa for ZnSb, does not work. However, if we keep in mind that the bulk modulus increases with the melting temperature [43] and that ZnSb has a larger formation energy than $Zn_4Sb_3$ [5,23], it becomes obvious why the bulk modulus of ZnSb is larger than the one of $Zn_4Sb_3$: indeed, the melting point of ZnSb is higher than the decomposition temperature of $Zn_4Sb_3$ [5]. This could also be one of the explanations why the elastic constants of ZnSb are smaller than in antimony-based skutterudites such as $CoSb_3$ for which B = 82-90 GPa and G = 57-61 GPa [44] and for which the peritectic temperature is higher than in ZnSb [5,45]. We note also that the bulk modulus of ZnSb is smaller than the one of $Mg_2Si$ whereas its shear modulus is larger than in $Mg_2Si$ (B = 70 GPa and G $\cong$ 21 GPa [46]). $Mg_2Si$ is a compound intensively studied for its thermoelectric properties during the last years. Therefore, the mechanical properties of ZnSb are comparable to those of other good thermoelectric materials for medium to high temperature applications. We find the same Poisson coefficient and the same $B_H/G_H$ ratio than experimentally. The $B_H/G_H$ ratio is slightly lower than 1.75, the limit between brittle and ductile behaviour as proposed by Pugh [47] and thus ZnSb is slightly brittle. This is not surprising since most of the semiconductors have such a behaviour. Indeed ductile materials are generally metallic [48] even though some metals can be brittle (see the example of $TiSi_2$ [41] or Iridium [49]). It is worth noting that we have recently found a transition from ductile behaviour (in the metallic $Tl_5Te_3$) to a slightly brittle behaviour in the small gap semiconductor $Tl_9XTe_6$ (with X = Sb or Bi) which is also a very good thermoelectric material [50]. The fact that the $B_H/G_H$ ratio is close to the boundary between ductile and brittle behaviour for ZnSb could thus be due to the small width of its band gap.

Now, we discuss about the consequences of our results on the anisotropy of the elastic properties. The case of one orthorhombic material, $TiSi_2$, has been discussed by Ravindran et al [41]. Following their procedure and using the same labelling, we have calculated the bulk modulus, the shear anisotropy factors A for the different directions ($A_1$:x, $A_2$:y and $A_3$:z), the anisotropies of the bulk modulus along the a axis and c axis with respect to the b axis ($A_{Ba}$ and $A_{Bc}$) and the percentage anisotropy in compressibility and shear ($A_B$ and $A_G$). The anisotropic Young's moduli for the different directions have also been determined ($E_i = 1/S_{ii}$, with i = 1, 2 or 3 and where $S_{ii}$ are the diagonal elements of the compliance matrix) and compared with the experimental values given in [40]. We have also calculated the shear moduli for the different directions [51]: $G_{xy} = C_{66}$, $G_{xz} = C_{55}$ and $G_{yz} = C_{44}$. Finally, we have determined the Poisson coefficient for the three directions x, y and z as follows [51]:

$v_{ij} = -S_{ij}/S_{ii}$, with i, j = 1, 2 or 3.

When the experimental counterparts of the calculated quantities were not directly available in the experimental work of Balaziuk et al [40], we have deduced them from the published experimental data.

In case of orthorhombic systems, the Cauchy relations are: $C_{12} = C_{66}$, $C_{13} = C_{55}$ and $C_{23} = C_{44}$ for respectively the x, y and z direction. In most of the solids, these relations are not fulfilled because the conditions are very restrictive and rely on the following assumptions [52, 53]:

- the absence of anharmonicity,
- the forces have to be central,
- the material has to be spatially isotropic.

In the real world, no material can fulfil all of these conditions even though some cubic materials can get close. In order to quantify the deviations from these conditions, it is usual to define the Cauchy pressure as $P^{Cauchy} = C_{12} - C_{44}$ in the case of cubic materials. In orthorhombic materials, we can define as well the Cauchy pressure for the three different directions: $P_x^{Cauchy} = C_{23} - C_{44}$, $P_y^{Cauchy} = C_{13} - C_{55}$ and $P_z^{Cauchy} = C_{12} - C_{66}$ (see table 4).

Most of the quantities measuring the anisotropy are either close to 1 ($A_1$, $A_2$, $A_3$, $A_{Ba}$, $A_{Bc}$) or very small ($A_B$, $A_G$): all these results indicate a relatively small anisotropy of the elastic properties of ZnSb, and are in good agreement with experiments.

From the results of the mechanical properties, we can extrapolate some conclusions on the character of the bonding in ZnSb. It is well known that purely ionic compounds like alkaline halides have higher Poisson coefficients than purely covalent compounds such as Si or C-diamond [48]. However, only recently this common affirmation has been confirmed more quantitatively by Ledbetter in a review paper [54]. Using Pauling's definition of the ionicity, Ledbetter was able to show strong evidence that the Poisson coefficient increases with the ionicity in body-centred cubic compounds with tetrahedral coordination. We can notably see in his review that ZnTe has a larger ionicity than GaSb and that the Poisson coefficient of the first compound is close to 0.3 compared to 0.247 in the second compound [54]. As can be seen in table 3, we have found a slightly larger powder-averaged Poisson coefficient in ZnSb (0.252) than the one of GaSb. This observation strongly supports the bonding analysis done by Benson et al [27] who have shown that the bonding characteristics of ZnSb are closer to those of GaSb than to those of the more ionic compound ZnTe, indicating that the bonding in ZnSb is much more covalent than ionic. This last result is also supported by recent spectroscopic data [55]. Our finding also indicates that the II-V family is very close to the III-V family concerning the chemical bonding but with a lower directional character since the bulk moduli are similar but the shear modulus of ZnSb is smaller than the one of GaSb.

The larger value of $C_{22}$ compared to $C_{11}$ and $C_{33}$ indicates that the bonding is strongest in the y direction. Therefore, the above data indicate that the y direction is the "hardest" direction for stretching processes but the "softest" direction for the shearing. Conversely, the x direction is the softest direction for stretching stresses but the hardest direction for shearing. The z direction has an intermediate behaviour with a bulk modulus slightly larger than the one along the x direction and a shear modulus slightly smaller than the one along the y direction. The Cauchy pressure can also give some information about the nature of the bonding. However, the situation is still not completely clear when discussing about the relation between the Cauchy pressure and the nature of the bonding. Indeed, if it has first been shown theoretically that the Cauchy pressure induced by an electron gas has to be positive, at least in the cubic case [56,57] (this has been confirmed in most of the cubic metals). Experimental works have also shown the possibility of a slightly or strongly negative Cauchy pressure in respectively Iridium [44,49] and rare-earth or actinide metals [53]. However, this is certainly related to the f electrons for these last metals, whereas for Iridium, it has been suggested that its unusual properties are related to the presence of strongly directional bonds [49]. It is well known also that semiconductors with diamond structure such as Si, C and Ge have a large and negative Cauchy pressure [58] which is related to the highly directional character of the bonds in these materials [56]. Thus, the suggestion done by Pettifor [56] that compounds with a positive Cauchy pressure tend to have metallic like bonds and that compounds with a negative Cauchy pressure have directional bonds is still qualitatively true. Concerning the ionic compounds, they can have both a large positive Cauchy pressure (AgI) or a large negative Cauchy pressure (MgO) [58,59]. However, it should be noted that in the case of more covalent intermetallic compounds, the Cauchy pressure has a strong tendency to decrease when ionicity decreases (see e.g. [54] and [58]). Therefore, from the above discussion, it seems that for cubic intermetallic tetrahedral covalent compounds, the smaller the Cauchy pressure is, the smaller the Poisson coefficient is. This behaviour occurs when the ionicity decreases and the Poisson coefficient and the Cauchy pressure are the smallest for highly directional bonds.

If this analysis of the Cauchy pressure can be extended to lower symmetry compounds as suggested elsewhere [60], this could be helpful for a better understanding of the nature of bonds in low symmetry complex materials such as ZnSb. However this analysis should be taken with care. In ZnSb, the $P_x^{Cauchy}$ is slightly negative and the $P_y^{Cauchy}$ is negative and relatively large, whereas the $P_z^{Cauchy}$ is positive and relatively large. It is interesting to note that the Cauchy pressure is slightly negative in GaSb (-3.1 GPa), whereas it is relatively large and positive in the more ionic ZnTe compound (9.7 GPa). Thus, the above comparisons confirm the conclusion drawn previously that ZnSb is a relatively covalent compound like GaSb. It is also worth noting that when the directions y and z (with the negative Cauchy pressure) are implied, one observes the lowest values of the Poisson coefficients. This observation confirms that the Cauchy pressure and the Poisson coefficients are related.

If, as discussed previously, the observation of a negative Cauchy pressure means that the bonds are more directional and less metallic, then the bonds in the y direction have a higher angular character than the bonds in the other directions. Conversely the bonds in the z direction have a more metallic character. It is interesting to see that the electrical conductivity $\sigma_z$ in the z direction is the largest, whereas the electrical conductivity $\sigma_y$ and hole mobility $\mu_{hy}$ in the y direction are the smallest [8,12]. This is consistent with the above discussion and can also be related to the electronic band structure shown in Fig. 2. Indeed, in the ΓY direction, the band gap is much larger than in the ΓX and ΓZ directions and in addition the band dispersion is smaller in the ΓY direction. Therefore, the mobility must be lower in the ΓY direction, in agreement with experiments, and the effective mass of the charge carriers along that direction must be larger. We note that in CdSb, which has a similar band structure than ZnSb, it was found that the $m^*_y$ effective mass is significantly larger than the other two effective masses [12]. Thus the anisotropy in the band structure can explain naturally the lower electrical conductivity and can also be related to the higher angular character in the y direction. However, it is more difficult to discuss the differences between the x and z directions solely from the band structure.

3.4 Lattice dynamics and thermal properties

As ZnSb crystallizes in a primitive orthorhombic structure with 8 ZnSb formula-units per primitive unit-cell, there are 48 different types of vibrational modes in the primitive unit-cell. At Γ point, these vibrational modes can be decomposed in irreducible modes as follows:

$$\Gamma_{vib} = \Gamma_{ac} + \Gamma_{opt} \quad (1)$$

with $\Gamma_{ac} = B_{1u} + B_{2u} + B_{3u}$ and $\Gamma_{opt} = 6 A_g + 6 B_{1g} + 6 B_{2g} + 6 B_{3g} + 6 A_u + 5 B_{1u} + 5 B_{2u} + 5 B_{3u}$.

As the $A_g$, $B_{1g}$, $B_{2g}$ and $B_{3g}$ are Raman active modes, there are 24 Raman modes and as the $B_{1u}$, $B_{2u}$ and $B_{3u}$ are infrared active modes, there are 15 infrared modes. The $A_u$ modes are silent modes.

In tables 5 and 6, we report respectively the Raman and infrared modes calculated at the Γ point. We also show the experimental results of Smirnov et al [61]. Our results agree reasonably well with these results for the infrared TO (transverse-optic) modes, with nevertheless a frequency downshift of about 5-10 %. This is essentially due to the overestimation of the cell volume in the relaxation calculations. Concerning the comparison with Raman experiments, the agreement is less good. Above 150 cm$^{-1}$ it is clear that the observation of broad peaks at about 170-180 cm$^{-1}$ in the calculated spectrum can be easily explained by the combination of two or three peaks expected in that energy range. We note the presence of a very low energy $A_g$ mode with a weak intensity at about 37.5 cm$^{-1}$ in the experiment which is absent in the calculations. However, the calculations predict a silent mode at about this energy. Also, we note the experimental observation of Raman modes at about 53 cm$^{-1}$ for the $B_{1g}$ and $B_{3g}$ symmetry and that no modes with these symmetries are predicted by the calculations at these energies. For these three cases, our best explanation is that the selection rules are probably relaxed by the presence of defects and particularly Zn vacancies (see next section). Recent Raman experiments on polycrystalline samples [42] confirm the presence of a low energy peak at 40 cm$^{-1}$. Other peaks are observed at 47 and 54 cm$^{-1}$ and also a broad and asymmetric peak at about 173 cm$^{-1}$: this can be easily deduced from our calculations. New polarized Raman experiments and inelastic neutron scattering experiments on ZnSb are needed to definitely conclude about this low energy Raman mode observed at about 40 cm$^{-1}$ (5 meV).

In figure 4, we show the phonon dispersion curves and the phonon density of states (DOS) obtained in our calculations for ZnSb. As can be seen from tables 5 and 6, the optic mode with the lowest energy has $A_u$ symmetry and is optically silent and therefore has a different symmetry than the acoustic modes close to the Brillouin zone center. However, the three following modes have $B_{2u}$, $A_u$ and $B_{2g}$ symmetry and the first mode is the most susceptible to mix with the acoustic mode that has the same symmetry. The acoustic modes at the Brillouin zone boundary and these four low energy optic modes are at the origin of the two low-energy peaks at 5.3 and 5.8 meV found in the phonon DOS.

Belash et al [62] in their inelastic neutron scattering (INS) study of amorphous Zn-Sb compounds have also studied a sample in which they have estimated to have 86 %-at. of orthorhombic ZnSb. In this sample, they see a clear peak in the INS spectrum at about 6.5 meV. Given that the instrumental resolution is about 2 meV in their experiment, their result agrees well with our calculations. Here again, we see that our calculations seems to underestimate slightly the energy of the lattice vibrations by about 10 %, as discussed above for the IR experiments. However it is necessary to make new INS experiments with both higher instrument resolution and broader energy range to confirm that and to permit a fine comparison between the phonon density of states and the lattice dynamics of ZnSb and $Zn_4Sb_3$ in order to understand the main reason of the very small thermal conductivity of this last compound (about 1 W/m.K) [2]. However, we note that the thermal conductivity of ZnSb is already relatively small (about 3 W/m.K) and even comparable with the ones of filled skutterudites [1, 4]. As in these last compounds, we observe the presence of low energy optic modes but in ZnSb the presence of these modes is just due to the complex crystal structure of orthorhombic ZnSb. Schweika et al [63] have done INS experiments on $Zn_4Sb_3$ and found that this compound has also a feature slightly above 5 meV, as in ZnSb. Recently, a more accurate position of this peak at about 7 meV was found by a combination of INS experiments and inelastic X-ray scattering on the $^{121}$Sb nucleus [64]. This is interesting, given the high similarity between the two crystalline structures and the presence of $Sb_2$ dimers in both cases as has been noted by several other authors [23,65] and also confirmed for ZnSb in section 3.2 where we find a strong bonding between Sb atoms. Since these $Sb_2$ dimers and the low energy modes at about 5-7 meV are present in both compounds, the very small thermal conductivity in $Zn_4Sb_3$ must have a different origin and is probably due to the large amount of disorder and defects in this last compound as suggested in the literature [1,4,6]. However, as can be seen in Fig. 5 where the partial phonon density of states is plotted, one sees that the low energy peaks imply equally the Zn atoms and the Sb atoms. This excludes the possibility that the low energy modes at about 5 meV are due to the rattling of the $Sb_2$ dimers at least for ZnSb. This is also confirmed by our preliminary mode analysis: clearly, the origin of the features at 5-7 meV in the Phonon DOS is much more complicated. This calls for further experimental investigation in both Zn-Sb compounds, but we think that our conclusion can also be extended to $Zn_4Sb_3$.

From the knowledge of the phonon density of states, we are able to calculate the thermodynamic properties and notably the vibrational heat capacity at constant volume $C_V$ and the vibrational entropy S. The results are reported in Figures 6 and 7 together with experimental results from [13] and [14]. Below 150 K, the calculations agree very well with the experiments and above 150 K, most of the difference can be accounted for by a contribution that grows linearly with the temperature (the $a_1T$ (fit of the data of [13]) and $a_2T$ (fit of the data of [14]) contributions to the experimental heat capacity curves in Fig. 7). In our calculations only the harmonic contribution at constant volume ($C_V$) has been considered which explains the disagreement with constant pressure ($C_P$) experiments above 150 K. Indeed other contributions to $C_P$ have to be taken into account:

$$C_P = C_V^{harm} + C_V^{QH} + C_V^{anharm} + C_V^{defects} \qquad (2)$$

where the first term, $C_V^{harm}$, is the harmonic contribution at constant volume and this is the contribution we have calculated; the second term, $C_V^{QH}$, is the contribution due to volume change, i. e. the quasi-harmonic contribution; the third term, $C_V^{anharm}$, is the purely anharmonic contribution; the last term, $C_V^{defect}$, is the contribution from the defects. The second term can be calculated from [66]:

$$C_V^{QH} = (B_M V \alpha_V^2)T \qquad (3)$$

where $B_M$ is the bulk modulus, V is the molar volume and $\alpha_V$ is the volume thermal expansion. The thermal expansion is necessary to evaluate this term but there is no report of its value. However, the lattice Grüneisen parameter, $\Gamma$, is related to the thermal expansion by the following relation [66]:

$$\Gamma = B_M \alpha_V V / C_V \qquad (4)$$

From our ab initio calculations it is possible to evaluate $\Gamma$ and thus $\alpha_V$ (because we have already determined $B_M$ and $C_V$) from the fit of the equation of states (EOS) with two different analytical formula. In the first case, we are using a relation implying dB/dP determined by fitting the energy vs volume curve with the Vinet EOS (see above). In that case we can use the crude Dingdale and Mcdonald approximation as follows [66]:

$$\Gamma^{DM} = -1/2 + (1/2)dB/dP \qquad (5)$$

This way, we find $\Gamma^{DM} = 2.175$. We can also use the semi-phenomenological EOS determined by Anton and Schmidt [67]:

$$E(V) = E_\infty + \frac{\beta V_0}{n+1}\left(\frac{V}{V_0}\right)^{n+1}\left(ln\left(\frac{V}{V_0}\right) - \frac{1}{n+1}\right) \qquad (6)$$

with β being the bulk modulus at equilibrium and n = 1/6 - $\Gamma^{AS}$. In that case, we find $\Gamma^{AS} = 1.99$ and the same value of the bulk modulus than when using the Vinet EOS. We note that these two values are significantly higher than for most of the thermoelectric materials and even than for $Zn_4Sb_3$ for which Caillat et al. found $\Gamma = 1.57$ [2] (in a recent work an even smaller value ($\Gamma = 1.35$) has been found [64]). In the case of filled skutterudites such as $RFe_4Sb_{12}$ that have roughly the same lattice thermal conductivity than ZnSb above 300 K, we have found previously $\Gamma = 1.5$ for the lattice Grüneisen parameter close to room temperature [68]. Thus, it seems that our calculated $\Gamma^{DM}$ and $\Gamma^{AS}$ are probably overestimated and is the upper bound for the lattice Grüneisen parameter. This is certainly due to the very crude approximation used to determine the Grüneisen parameter: the relatively large errors in the determination of the fitting parameters in the fit of the EOS, the overestimation of the lattice parameters and the anisotropy of the Grüneisen parameter that we have not taken into account. Therefore, we will overestimate the thermal expansion in Eq.4 as well as $C_V^{QH}$ in Eq.3. Since $\Gamma^{AS}$ has a smaller value and closer to the one found in other thermoelectric materials, we think it is more realistic than $\Gamma^{DM}$ and we will therefore use $\Gamma^{AS}$ in the following. Finally using the calculated heat capacity at 300 K together with the bulk modulus obtained from the equation of state, we find $\alpha_V = 6.5*10^{-5}$ $K^{-1}$ from Eq.4. Experimental data are not available for the thermal expansion of ZnSb to permit a comparison. However, we note that the above value is much larger than the value found for $Zn_4Sb_3$ ($\alpha_V = 3*10^{-5}$ $K^{-1}$) by Nakamoto et al [69] and this is again an indication that we probably overestimate the thermal expansion by determining it using Eqs. 4 and 5. If we try to calculate the value of $C_V^{QH}$ in using the thermal expansion calculated from Eq.4, we find a high temperature slope of 0.053 which is too small compared to the slope extracted from the data of ref. 13 (the slope is 0.08) and almost three times smaller than the slope obtained from ref. 14 (the slope is 0.15). Since the value of the thermal expansion we have used is an upper boundary of the thermal expansion, as discussed above, we can infer that in all cases there must be an additional linear contribution to the heat capacity above 150 K.

We have tried to estimate the contribution from the defects (in our case the vacancies, see in the last section) using the following semi-phenomenological formula that was found to work well in the case of aluminium [70]:

$$C_V^{defects} = Nk_b\, e^{\Delta S/k_b} (E/k_bT)^2\, e^{E/k_bT} \qquad (7)$$

where E and $\Delta S$ are respectively the formation energy and the formation entropy of a zinc vacancy. We consider only this defect as it is the most stable one (see following section). Even by considering unreasonably high values for $\Delta S$, we find that the contribution of $C_V^{defects}$ is negligible at room temperature and below.

Having eliminated the other possibilities, we have to examine the case of the anharmonic contribution which is given as follows for the two smallest order terms [70-72]:

$$C_V^{anharm} = -T\delta^2(F_3+F_4)/\delta T^2 = 3RBT \qquad (8)$$

where $F_3$ is the first order cubic term of the free energy and $F_4$ is the first order quartic term of the free energy. Because both terms have a $T^2$ temperature dependence, their contribution to the specific heat is linear with the temperature. At least in metals, the $F_3$ term is generally negative whereas the $F_4$ term is generally positive. Therefore, when one finds a large positive linear term due to anharmonicity in the specific heat above room temperature, after subtraction of the quasi-harmonic term, this means that the cubic term $F_3$ is larger than the quartic term $F_4$. We believe that this is the case here in ZnSb and that the observation of a large anharmonic contribution to the heat capacity has to be related to its low thermal conductivity. However, in order to estimate with reliability the value of the anharmonic term B, it is necessary to have much more accurate heat capacity data and also to perform thermal expansion experiments in order to estimate both $\Gamma$ and $C_V^{QH}$.

For the entropy, the agreement is better than for the heat capacity because the calculated harmonic entropy alone reproduces very well the experimental data. It is interesting to note that if we plot our calculated heat capacity as $C_V/T^3$ vs T, we find a maximum at about 13 K (see inset in Fig. 6). This is what would be expected in the case of a simple Einstein model with an Einstein energy of about 5.4 meV. This energy corresponds to the two low energy peaks found in the phonon DOS at about 5.5 meV, as discussed above. Therefore, in the case of ZnSb, the observation of an Einstein-like behaviour in the heat capacity has nothing to do with the presence of an Einstein mode. Note that a similar value for an Einstein mode has been found for $Zn_4Sb_3$ in fitting the experimental heat capacity with combined Debye and Einstein models [64]. This is probably due to the similarities between the phonon densities of states of ZnSb and $Zn_4Sb_3$ at low energy which are related to the proximity of the crystal structure of the two compounds as discussed previously. From the knowledge of the phonon density of states, it is possible to calculate the tensor of the Atomic Displacement Parameters (ADP) $U_{ij}$ (i, j = x, y, z) of the different atom types in a given crystal structure [35]. We show the averaged $U_{ij}$ for the zinc and antimony atoms in the ZnSb structure at 300 K in table 7. It is worth noting that the diagonal elements of the ADP tensor are the same for all the atoms of a given type, whereas this is not the case for the three off-diagonal elements $U_{zx}$, $U_{yz}$ and $U_{xy}$. Indeed, exactly as in the case of the electronic properties, the ADP off-diagonal elements are the same only for pairs of Zn and Sb atoms.

How can we interpret these results concerning the off-diagonal elements of the ADP ? The observation of different off-diagonal elements of the ADP and the observation of different electronic density of states for atoms that are in equivalent Wyckoff positions could mean that in fact the symmetry of our ZnSb sample is lower than orthorhombic and that there could be a very small internal distortion that lowers the symmetry. Since the atoms are located in sites of very low symmetry (in fact they are located in the general Wyckoff positions of the Pbca space group which explains why off-diagonal elements of the ADP can be non zero), the distortion should probably lead to a triclinic symmetry. We cannot exclude a possible calculation artefact but as mentioned earlier we have carefully checked our calculations and we systematically find that the atoms go by pairs that have different properties: ADPs, electronic and vibrational properties (we find indeed also small differences in the phonon density of states (not shown) for the different atomic pairs). From an experimental point of view, this possibility could be checked only with high precision diffraction experiments on single-crystals. Actually, there is only one experimental work in which the ADPs have been determined [73]. This was done from a Rietveld refinement of an X-ray pattern of a powder sample in which only the isotropic ADPs $B_{iso}$ ($B_{iso}=8\pi^2 U_{eq}$ with $U_{eq} = 1/3 \Sigma U_{ii}$) have been determined and assuming full occupancy of the atom sites. Therefore this experimental result has to be taken with caution when comparing with our calculations. At room temperature, Mozharivskyj et al. have found $B_{iso}$= 1.5(1) Å$^2$ for the Zn atoms and $B_{iso}$= 1.28(5) Å$^2$ for the Sb atoms [73], whereas we find $B_{iso}$= 1.58 Å$^2$ for the Zn atoms and $B_{iso}$= 1.02 Å$^2$ for the Sb atoms. This is a reasonable agreement between experiments and simulations taking into account the above comments. Better experiments such as high resolution neutron diffraction experiment are needed in order to make an in depth comparison with our results, especially concerning the anisotropy of the ADP parameters.

To further explore the properties of the phonons in ZnSb and to compare with the ones of $Zn_4Sb_3$, we have calculated the Debye temperature ($\Theta_D$) from the averaged sound velocity ($v_D$) obtained with our calculations and also with the experimental data of ref. 40 by using the following equation:

$$\Theta_D = \frac{h}{k_B}\left(\frac{3}{4\pi V_a}\right)^{1/3} v_D \qquad (9)$$

where $h$ and $k_B$ are, respectively, Planck's and Boltzmann's constant and $V_a$ is the atomic volume. The average sound velocity in polycrystalline systems, $v_m$, are evaluated by

$$\frac{1}{v_D^3} = \frac{1}{3}\left(\frac{1}{v_l^3} + \frac{2}{v_t^3}\right) \qquad (10)$$

where $v_l$ and $v_t$ are the mean longitudinal and transverse sound velocities, which can be related to the shear and bulk moduli:

$$v_l = \left(\frac{3B+4G}{3\rho}\right)^{1/2} \text{ and } v_t = \left(\frac{G}{\rho}\right)^{1/2} \qquad (11)$$

In this equation, we are using the powder-averaged shear and bulk moduli, $G_H$ and $B_H$, as determined above with the Hill's procedure. As can be seen in Ravindran's work [41], this procedure gives correct values in the case of the orthorhombic structure.

We have also calculated the Debye temperature from the zero point vibrational energy $E_0$ per unit-cell by using the following relation [66]:

$$\Theta_D^0 = (9E_0/8nR) \qquad (12)$$

where R is the gas constant and n the number of atoms per unit-cell.

From our calculated heat capacity, we can also calculate the Debye temperature $\Theta_D^C$ from the $T^3$ dependence of the heat capacity at low temperatures and using [74]:

$$\Theta_D^C = (12\pi^4 R/5\beta)^{1/3} \qquad (13)$$

where β is the coefficient of the $T^3$ term of the heat capacity at low temperature. We find a value of 209.3 K. This value is smaller than the one obtained from the elastic constants (236.1 K) and much smaller than the one obtained from the zero point vibrational energy $E_0$ = 0.0209eV/at. (i. e. $\Theta_D^0$ = 272.8 K). Ravindran et al [41] have already observed the same tendency between $\Theta_D^C$ and $\Theta_D$ obtained from the elastic constants in the case of orthorhombic $TiSi_2$. These differences are not unexpected since the values of the Debye temperatures obtained from different definitions/experiments are expected to be different, although close (a discussion of this aspect is beyond the scope of the present article and the reader can find more information about this observation in the review of Barron et al [66]). The important point is to compare experiments and calculations of the same Debye temperature, as is done below in the case of the one calculated from the elastic constants.

The experimental values from ref. 40 ($\Theta_D$ = 253 K) and ref. 14 ($\Theta_D$ = 208 K) are respectively larger and smaller than the calculated value obtained from the elastic constants ($\Theta_D$ = 236.1 K). As discussed previously, the value found in ref. 40 is the most reliable. We would like to stress that our calculations are done at 0 K, while the Debye temperature $\Theta_D$ is obtained from ultrasonic experiments performed at room temperature and this can explain a part of the difference between calculations and experiments. However, as we have seen above, the main origin of the differences between our calculations and the experiments is the larger volume found in our relaxation calculations which decreases the value of the sound velocity and also of the Debye temperature.

In the next step, we aim to estimate the thermal conductivity, $\kappa$, using a very simple model considering only Umklapp scattering in order to see if this mechanism can be the dominant mechanism of the phonons. We use the following formula employed in the case of $Zn_4Sb_3$ by Caillat et al [2]:

$$\kappa = A\, M_{at}\, (V_{at})^{1/3}\, \Theta_D^2 / (n^{1/3} \Gamma)^2 \qquad (14)$$

where $M_{at}$ is the average atomic mass, $V_{at}$ is the atomic volume, $\Theta_D$ is the Debye temperature, A is a constant equal to $3.17 \times 10^7$ s$^{-3}$K$^{-3}$, n is the number of atoms in the unit-cell and $\Gamma$ is the Grüneisen parameter. If we want to determine the thermal conductivity from our calculations, we need the Grüneisen parameter. Thus, we use the Grüneisen parameter estimated from Eq. 6 with the fit of the EOS with the Anton-Schmidt formula for the reasons discussed previously.

Using this value of $\Gamma^{AS}$ = 1.99 and the Debye temperature calculated from the elastic constants, we find $\kappa$ = 1.93 W/m.K. If we use the experimental volume and the Debye temperature derived from the experimental elastic constants [40], we find $\kappa$ = 2.21 W/m.K. These values are about two times smaller than the measured values ($\kappa$ = 3-4 W/m.K) [4, 8, 10]. However, as discussed previously, we have used an overestimated value of the Grüneisen parameter determined in an unusual manner and therefore we need to have a more reliable determination of the Grüneisen parameter for ZnSb. If we use the same Grüneisen parameter than for $Zn_4Sb_3$ (i. e. $\Gamma$ =1.57 [2] or 1.35 [65]), we find $\kappa$ = 3.1 or 4.19 W/m.K if the other parameters are the calculated parameters and we find $\kappa$ = 3.51 or 4.75 W/m.K if we use the experimental parameters. As can be seen, a better agreement with the experimental thermal conductivity is found in the last cases and this confirms that $\Gamma$ is probably closer to 1.5 than to 2. Therefore, we think that it is necessary to determine with a good accuracy the Grüneisen parameter of ZnSb in order to definitely conclude. Nevertheless, we note that we find the good order of magnitude when we calculate the thermal conductivity using the Umklapp mechanism and a Grüneisen parameter similar to the one of $Zn_4Sb_3$ and this shows that this mechanism is a good candidate to explain the relatively low thermal conductivity of ZnSb. Finally, it is worth noting that from the comparison between Balazyuk's ultrasonic results for ZnSb and recent ultrasonic experiments for $Zn_4Sb_3$ [75] (see Table 8), the sound velocity and the Debye temperature of ZnSb and $Zn_4Sb_3$ are very similar. This observation together with the very similar phonon density of states below 10 meV in the two materials suggests the following reasons to explain the lower thermal conductivity of $Zn_4Sb_3$. Firstly, if the above suggested Umklapp mechanism was the main mechanism of phonon scattering, it would mean that the main reason for the lower thermal conductivity in $Zn_4Sb_3$ is the larger number of atoms in its unit-cell. However the disorder and the number of defects are much higher in $Zn_4Sb_3$ than in ZnSb [1, 4, 73]. Since this disorder and these defects induce an additional scattering mechanism of the heat-carrying phonons in $Zn_4Sb_3$ compared to ZnSb, this provides a natural explanation of the lower thermal conductivity observed in $Zn_4Sb_3$.

This last proposal also explains naturally why nanostructuring ZnSb is very efficient to reduce the thermal conductivity of ZnSb towards values observed in $Zn_4Sb_3$ as shown recently by several groups [18,19]. Indeed, these groups were able to obtain thermal conductivities as low as 1.4-2 W/m.K with a grain size of a few tens nanometers. There is however still a potential to further decrease the lattice thermal conductivity of ZnSb by decreasing further the grain size and/or increasing the point defect scattering because these values are still three to four times larger than the minimum lattice thermal conductivity $\kappa_{min}$ that is about 0.51 W/m.K at room temperature ($\kappa_{min} = 1/3\, C_V\, v_m\, d$ with $d$ = 2.732 Å being the mean calculated interatomic distance [76]).

## 3.5 Defect stability

Defects such as vacancies, antisites and interstitial atoms were inserted in the 2x2x2 supercell. The formation energy of a particular defect in ZnSb in eV/defect can be calculated from the following equation:

$$E_D = \frac{\Delta H_D(ZnSb^D) - \Delta H_0(ZnSb)}{x_D} \qquad (15)$$

where $\Delta H_D(ZnSb^D)$, $\Delta H_0(ZnSb)$ and $x_D$ are respectively :

- the formation enthalpy calculated (in eV/atom) for the 2x2x2 supercell of ZnSb containing the corresponding defect

- the formation enthalpy calculated (in eV/atom) for the 2x2x2 supercell of ZnSb without the defect

- the concentration of the defects in the 2x2x2 supercell of ZnSb

The first two values are calculated using equation (1).

With this procedure we have determined the stability of the different types of intrinsic defects in ZnSb. Since ZnSb is a p-type intrinsic semiconductor, its most stable defects should be either zinc vacancies $V^{Zn}$, antisites $Sb^{Zn}$ (Sb on a zinc site) or interstial Sb, $I^{Sb}$. The results of our calculations are reported in table 6. We find that the most stable defect is indeed $V^{Zn}$. This result agrees well with Calphad assessment of the ternary Zn-Cd-Sb phase diagram recently done in our group [5, 77] and by experimental observations done by Mozharivskyj et al [73]. We find that the most stable (Zn,Cd)Sb phases are rich in Sb and have Zn/Cd vacancies. The domain of existence of the Sb-rich (Zn,Cd)Sb phases is found to be larger when the temperature is increased. To confirm that Zn vacancies induce a p-type doping in ZnSb, we have calculated the corresponding electronic density of states. As can be seen in Fig. 9, the effect of 1 vacancy in the Zn sublattice (in a supercell containing 127 atoms) is to down-shift the Fermi level 0.32 eV below the Fermi level of the perfect supercell, confirming thus that the presence of $V^{Zn}$ induces p-doping in ZnSb. Note that from the above mentioned Calphad study, it seems that this conclusion can be extended to all the compounds with $Zn_{1-x}Cd_xSb$ composition. We therefore claim that the most stable defects in CdSb are Cd vacancies and this is confirmed experimentally [78]. In the search of optimized thermoelectric properties of ZnSb (and related compounds) by doping, one has to consider the possible effect of these vacancies that can compensate the effect of the inserted impurities. This is particularly true for donor type impurities which can explain the difficulty to design n-type ZnSb based materials.

## 4. Conclusion

Our first principles calculations of the physical properties of orthorhombic ZnSb are in good agreement with previous calculations done recently in the literature. We show that electronically not all Zn (Sb) atoms are strictly equivalent. This is confirmed by the vibrational properties and the atomic displacement parameters. The analysis of the Poisson coefficient and the Cauchy pressure is in agreement with the relatively strong covalent character of the bonding in ZnSb. We find that the elastic properties have a relatively small anisotropy in good agreement with experiments and that the bonds along the b-direction are the strongest.

The elastic constants are comparable to those of other good high temperature thermoelectric materials and therefore from a mechanical point of view, ZnSb can be used for high temperature thermoelectric applications. We have found the presence of low energy vibrational modes in the phonon density of states whose interactions with the acoustic phonons could explain the relatively low thermal conductivity of ZnSb. The thermal conductivity could be further decreased to values observed in $Zn_4Sb_3$ by nanostructuring ZnSb, as some first results have shown [18, 19]. Finally, we have shown that the most stable defect in orthorhombic ZnSb is the Zinc vacancy which explains naturally why it is intrinsically p-doped. This must be taken into account when ZnSb is doped, especially for n-doping, in order to avoid undesirable compensation effects. The "good" physical properties listed above added to its better stability and its ability to be n-doped in contrast to $Zn_4Sb_3$ explain why orthorhombic ZnSb is a promising thermoelectric material even though still underestimated.

**Acknowledgments:** X. T. thanks the National Natural Science Foundation of China (11047031) and the Guangxi Natural Science Foundation (2011GXNSFC018003). We thank the computer centers CINES and HPC@LR in Montpellier for their support. We acknowledge the financial support of Hutchinson Industries, Total S.A. and of the CNRS through the "Programme Interdisciplinaire Energie".

References

[1] G.J. Snyder, E.S. Toberer, *Nature Mater.* **7**, 105 (2008)

[2] T. Caillat et al, *J. Phys. Chem. Solids* **58**, 1119 (1997)

[3] V. Izard, M.-C. Record, J. C. Tedenac, S. G. Fries, *Calphad* **25**, 567 (2001)

[4] G. J. Snyder et al, *Nat. Mater.* **3**, 458 (2004) and ref. therein.

[5] Y. Liu, J. C. Tedenac, Calphad **33**, 684 (2009) and ref. therein.

[6] See e. g. G. S. Pomrehn, E. S. Toberer, G. J. Snyder, A. Van de Walle, *Phys. Rev. B* **83**, 094106 (2011)

[7] H. Kmoiya et al, *Phys. Rev.* **133**, A1679 (1964)

[8] P. J. Shaver, J. Blair, *Phys. Rev.* **141**, 649 (1966)

[9] A. Abouzeid, G. Schneide, *Phys. Stat. Sol. (a)* **6**, K101 (1971)

[10] I. V. Dakhovskii et al, *Sov. Phys. Semicond.* **15**, 423 (1981) and ref. therein

[11] A. Abouzeid, G. Schneide, *Zeitschrift für Naturforsch. A* **30**, 381 (1975)

[12] For a review, see E. K. Arushanov, *Prog. Crystal Growth and Charact.* **13**, 1 (1986)

[13] K. M. Mamedova, A. Yu. Dzhangirov, O. I. Dzhafarov, V. N. Kostryukov, *Rus. J. Phys. Chem.* **49**, 1635 (1975)

[14] G. N. Danilenko, V. Ya. Shevchenko, S. F. Marenkin, M. Kh. Karapet'yants, *Inorg. Mat.* **14**, 486 (1978)

[15] T. A. Stolyarova, *Russian Metall.* (6) 61 (1979)

[16] V. A. Goncharuk, G. M. Lukashenko, *Rus. J. Phys. Chem.* **62**, 834 (1988)

[17] V. A. Goncharuk, G. M. Lukashenko, *Inorg. Mat.* **25**, 1621 (1989)

[18] P. H. M. Böttger et al, *J. Electron. Mat.* **39**, 1583 (2010) ; P. H. M. Böttger et al, Proc. ECT 2010, p. 61.

[19] C. Okamura et al, *Mater. Trans.* **51**, 860 (2010)

[20] T. Ueda et al, *Mater. Trans.* **50**, 2573 (2009)

[21] K.-W. Jang, H.-J. Oh, I.-H. Kim, J.-I. Lee, *Electron. Mat. Lett.* **6**, 193 (2010)

[22] Y. Yamada, *Phys. Stat. Sol. (b)* **85**, 723 (1978)

[23] A. S. Mikhaylushkin, J. Nylen, U. Haussermann, *Chem. Eur. J.* **11**, 4912 (2005)

[24] P. Boulet, M.-C. Record, *Solid State Science* **12**, 26 (2010)

[25] J.-H. Zhao, E.-J. Han, T.-M. Liu, W. Zeng, *Int. J. Mol. Sci.* **12**, 3162 (2011)

[26] L. Bjerg, G. H. K. Madsen, B. B. Iversen, *Chem. Mat.* **23**, 3907 (2011)

[27] D. Benson, O. F. Sankey, U. Häussermann, *Phys. Rev. B* **84**, 125211 (2011)

[28] P. E. Blöchl, *Phys. Rev. B* **50**, 17953 (1994)

[29] G. Kresse, D. Joubert, *Phys. Rev. B* **59**, 1758 (1999)

[30] G. Kresse, J. Furthmuller, *Phys. Rev. B* **54**, 11169 (1996)

[31] J. P. Perdew, K. Burke, M. Ernzerhof, *Phys. Rev. Lett.* **77**, 3865 (1996)

[32] H. J. Monkhorst, J. D. Pack, *Phys. Rev. B* **13**, 5188 (1976)

[33] P. E. Blöchl, O. Jepsen, O. K. Andersen, *Phys. Rev. B* **49**, 16223 (1994)

[34] P. Vinet, J. H. Rose, J. Ferrante, J. R. Smith., *J Phys: Condens Matter* **1**, 1941 (1989)

[35] K. Parlinski, Software Phonon, Cracow (2010) ; K. Parlinski, Z. Q. Li, Y. Kawazoe, *Phys. Rev. Lett.* **78**, 4063 (1997) ; K. Parlinski, *J. Phys.: Cond. Mat.* **78**, 012009 (2007) and ref. therein.

[36] M. Methfessel, A.T. Paxton, *Phys. Rev. B* **40**, 3616 (1989)

[37] X. M. Tao, P. Jund, C. Colinet, J. C. Tedenac, *Phys Rev B* **80**, 104103 (2009)

[38] F. L. Carter, R. Mazelsky, J. Phys. Chem. Solids **25**, 571 (1964)

[39] J. Hafner, *J. Comput. Chem.* **29**, 2044 (2008)

[40] V. N. Balazyuk, A. I. Eremenko, N. D. Raransky, *Func. Mat.* **15**, 343 (2008).

[41] P. Ravindran et al, *J. Appl. Phys.* **84**, 4891 (1998)

[42] D. M. Triches, S. M. Souza, J. C. de Lima, T. A. Grandi, C. E. M. Campos, A. Polian, J. P. Itié, F. Baudelet, J. C. Chervin, *J. Appl. Phys.* **106**, 013509 (2009)

[43] K. Moriguchi, M. Igarashi, *Phys. Rev. B* **74**, 024111 (2006)

[44] L. Zhang et al, *Mat. Science Eng. B* **170**, 26 (2010)

[45] See e. g. Y. Zhang, C. Li, Z. Du, C. Guo, J. C. Tedenac, Calphad **33**, 405 (2009)

[46] S. Ganeshan, S. L. Shang, Y. Wang, Z. K. Liu, *J. Alloys Compds* **498**, 191 (2010)

[47] S.F. Pugh, *Philos. Mag.* 45, 823 (1954)

[48] J. Haines, J. M. Léger and G. Bocquillon, *Ann. Rev. Mater. Res.* **31**, 1 (2001)

[49] S. Kamran, K. Chen, L. Chen, *Phys. Rev. B* **79**, 024106 (2009)

[50] X. Tao, P. Jund, R. Viennois, J.-C. Tedenac, *J. Phys. Chem. A* **115**, 8761 (2011).

[51] J. Lemaitre, J.-L. Chaboche, A. Benallal, R. Desmorat, *Mécanique des matériaux solides*, Dunod (Paris), 2009 (in french)

[52] D. J. Quesnel, D. S. Rimai, L. P. DeMejo, *Phys. Rev. B* **48**, 6795 (1993)

[53] H. Ledbetter, A. Migliori, *Phys. Stat. Sol. (b)* **245**, 44 (2008)

[54] H. Ledbetter, S. Kim in *Handbook of Elastic Properties of Solids, Liquids, and Gases*, edited by Levy,

Bass, and Stern, Volume II, ch. 16, p. 281.

[55] P. H. M. Böttger, S. Diplas, E. Flage-Larsen, O. Prytz, T. G. Finstad, *J. Phys. : Cond. Mat.* **23**, 265502 (2011)

[56] D. G. Pettifor, *Mater. Science Technol.* **8**, 345 (1992)

[57] R. A. Johnson, *Phys. Rev. B* **37**, 3924 (1988)

[58] H. Ledbetter, S. Kim in *Handbook of Elastic Properties of Solids, Liquids, and Gases*, edited by Levy, Bass, and Stern, Volume II, ch. 3, p. 57.

[59] R. A. Bartels, P. R. Son, *J. Phys. Chem. Solids* **33**, 749 (1972)

[60] S. Ganeshan, S. L. Shang, H. Zhang, Y. Wang, M. Mantina, Z. K. Liu, *Intermet.* **17**, 313 (2009)

[61] D. V. Smirnov, D. V. Mashivets, S. Pasquier, J. Leotin, P. Puech, G. Landa, Yu. V. Roznovan, *Semicond. Sci. Technol.* **9**, 333 (1994)

[62] I. T. Belash, O. I. Barbakov, A. I. Kolesnikov, E. G. Ponyatovskii, M. Prager, *Solid State Com.* **78**, 331 (1991).

[63] W. Schweika, R. P. Hermann, M. Prager, J. Persson, V. Keppens, *Phys. Rev. Lett.* **99**, 125501 (2007).

[64] A. Möchel, I. Segueev, H.-C. Wille, F. Juranyi, H. Schober, W. Schweika, S. R. Brown, S. M. Kauzlarich, R. P. Hermann, *Phys. Rev. B* **84**, 184303 (2011) and ref. therein.

[65] U. Häussermann, A. S. Mikhaylishkin, *Dalton Trans.* **39**, 1036 (2010)

[66] T. H. K. Barron, J. G. Collins, G. K. White, *Adv. Phys.* **29**, 609 (1980).

[67] H. Anton, P. C. Schmidt, *Intermet.* **5**, 449 (1997)

[68] R. Viennois, S. Charar, D. Ravot, P. Haen, A. Mauger, A. Bentien, S. Paschen, F. Steglich, *Eur. Phys. J. B* **46**, 257 (2005)

[69] G. Nakamoto, K. Kinoshita, M. Kurisu, *J. Alloys Compds* **436**, 65 (2007).

[70] R. C. Shukla, C. A. Plint, *Int. J. Thermophys.* **1**, 299 (1980).

[71] R. C. Shukla, C. A. Plint, *Int. J. Thermophys.* **1**, 73 (1980).

[72] G. Leibried, W. Ludwig, *Solid State Phys.* **12**, 275 (1961).

[73] Y. Mozharivskyj, A. O. Pecharsky, S. Bud'ko, G. J. Miller, *Chem. Mater.* **16**, 1580 (2004).

[74] See e. g. C. Kittel, *Introduction to Solid State Physics*, Wiley; 7th edition (1995)

[75] S. Bhattacharya, R. P. Hermann, V. Keppens, T. M. Tritt, G. J. Snyder, *Phys. Rev. B* **74**, 134108 (2006).

[76] V. K. Zaitsev, M. I. Fedorov, in *Thermoelectrics Handbook-Macro to Nano*, edited by D. M. Rowe., CRC Press, Taylor & Francis, Boca Raton, ch. 6, 2006.

[77] Y. Liu, J. C. Tedenac (unpublished results).

[78] B. N. Gritsyuk, A. I. Rarenko, A. V. Sirota, D. D. Khalameida, *Semicond.* **27**, 852 (1993) and references therein.

Table Captions

Table 1 Calculated and experimental lattice constants and formation enthalpies of ZnSb

Table 2 Irreducible strains, distortions and corresponding elastic constants of an orthorhombic system

Table 3 Calculated elastic constants, bulk modulus, shear modulus, Young's modulus, Poisson's ratio and $B_H/G_H$ of ZnSb compared with experimental results[14,40]

Table 4 Calculated and experimental anisotropic elastic constants[40]

Table 5 Energies of the calculated Raman-active vibrational modes compared with the experimental values[61] (in wave-number)

Table 6 Energies of the calculated Infrared-active and silent vibrational modes compared with the experimental values of the infrared vibrational modes[61] (in wave-number)

Table 7 Averaged anisotropic atomic displacement parameters $U_{ij} = \langle u_i u_j \rangle$ (i, j = x, y, z) of zinc and antimony atoms for the ZnSb structure.

Table 8 Calculated longitudinal, transverse, average sound velocities, and Debye temperature of ZnSb compared to the experimental values of ZnSb and $Zn_4Sb_3$[2, 40, 74]

Table 9 Calculated formation energy of the native point defects of ZnSb

Figure Captions

Fig.1 ZnSb structure showing the $Zn_2Sb_2$ rhomboid rings projected along the 3 directions of the orthorhombic cell

Fig. 2 Electronic band structure of ZnSb

Fig. 3 DOS of the four inequivalent Zn and Sb atoms in the unit cell

Fig. 4 Phonon dispersion curves (different colors correspond to different symmetries) and total phonon density of states of ZnSb

Fig. 5 Partial phonon density of states for Sb and Zn atoms of ZnSb

Fig. 6 Calculated heat capacity compared to experimental results[13,14]. An additional term proportional to the temperature is needed to reproduce the experiments (see text for details)

Inset : plot of the calculated $C_V/T^3$ and experimental $C_P/T^3$ vs T

Fig. 7 Calculated entropy compared to experimental results (full squares [13], hollow circles [14])

Fig. 8 Total electronic DOS of $Zn_{64}Sb_{64}$ (supercell without defect) and $Zn_{63}Sb_{64}$ (supercell with one Zn vacancy)

Table 1

|  | Lattice parameters (Å) | | | Formation enthalpy ($\Delta H$) eV/atom | Atomic positions/Remarks |
| --- | --- | --- | --- | --- | --- |
|  | a | b | c |  |  |
| GGA | 6.2808 | 7.8246 | 8.2293 | -0.0346 | Zn (0.0407,0.1063,0.1271) Sb (0.3584,0.417,0.3907) |
| LDA | 6.1086 | 7.5817 | 7.9859 | -0.002 | Zn (0.039,0.1034,0.1261) Sb (0.3568,0.4148,0.3893) |
| Experiment | 6.2016 | 7.7416 | 8.0995 |  | Zn (0.0414,0.1128,0.132) Sb (0.358,0.4188,0.3923)[38] |
|  |  |  |  | -(0.0814-0.0825) | At 670 K (galvanic cell and emf method) [16,17] |
|  |  |  |  | -0.0665 | At 298 K (vacuum block calorimetry)[15] |
|  |  |  |  | -0.0778 | (calorimetry, ref. 7 in [17]) |
|  |  |  |  | -0.0774 | (dissociation pressure method, ref. 8 in [17]) |
|  |  |  |  | -0.0934 | Ref. 60 in [12] |

Table 2

| structure | strain | distortion | $\Delta E/V$ [2$^{nd}$ order in δ ] |
| --- | --- | --- | --- |
| orthorhombic | $\varepsilon_1$ | $e_1=\delta$ | $C_{11}\delta^2/2$ |
|  | $\varepsilon_2$ | $e_2=\delta$ | $C_{22}\delta^2/2$ |
|  | $\varepsilon_3$ | $e_3=\delta$ | $C_{33}\delta^2/2$ |
|  | $\varepsilon_4$ | $e_4=2\delta$ | $2C_{44}\delta^2$ |
|  | $\varepsilon_5$ | $e_5=2\delta$ | $2C_{55}\delta^2$ |
|  | $\varepsilon_6$ | $e_6=2\delta$ | $2C_{66}\delta^2$ |
|  | $\varepsilon_7$ | $e_1=(1-\delta^2)^{-1/3}(1+\delta)-1$ $e_2=(1-\delta^2)^{-1/3}(1-\delta)-1$ $e_3=(1-\delta^2)^{-1/3}-1$ | $(C_{11}+C_{22}-2C_{12})\delta^2/2$ |
|  | $\varepsilon_8$ | $e_1=(1-\delta^2)^{-1/3}(1+\delta)-1$ $e_2=(1-\delta^2)^{-1/3}-1$ $e_3=(1-\delta^2)^{-1/3}(1-\delta)-1$ | $(C_{11}+C_{33}-2C_{13})\delta^2/2$ |
|  | $\varepsilon_9$ | $e_1=(1-\delta^2)^{-1/3}-1$ $e_2=(1-\delta^2)^{-1/3}(1+\delta)-1$ $e_3=(1-\delta^2)^{-1/3}(1-\delta)-1$ | $(C_{22}+C_{33}-2C_{23})\delta^2/2$ |

Table 3

|  | GGA | Experiment (T=300K) |
|---|---|---|
| $C_{11}$ (GPa) | 80.2 | 92.4[40] |
| $C_{22}$ (GPa) | 93.3 | 103[40] |
| $C_{33}$ (GPa) | 84.4 | 93.6[40] |
| $C_{12}$ (GPa) | 29.5 | 32.9[40] |
| $C_{23}$ (GPa) | 26 | 31.1[40] |
| $C_{13}$ (GPa) | 29 | 38.4[40] |
| $C_{44}$ (GPa) | 18.5 | 21.6[40] |
| $C_{55}$ (GPa) | 37.6 | 46.3[40] |
| $C_{66}$ (GPa) | 30.2 | 36[40] |
| Bulk Modulus $B_H$ (GPa) | 47.35 | 54.75[40] |
| Shear Modulus $G_H$ (GPa) | 28.03 | 32.16[40], 45[14] |
| Young's Modulus E (GPa) | 70.2 | 80.6[40], 105[14] |
| Poisson's ratio ν | 0.252 | 0.253[40] |
| $B_H/G_H$ | 1.69 | 1.7[40] |

Table 4

|  | GGA | Experiment[40] (T=300K) |
|---|---|---|
| $B_a$ (GPa) | 133.4 | 162.4 |
| $B_b$ (GPa) | 158.6 | 170.8 |
| $B_c$ (GPa) | 136.2 | 160.9 |
| $G_{yz}$ (GPa) | 37.6 | 46.3 |
| $G_{xz}$ (GPa) | 18.5 | 21.6 |
| $G_{xy}$ (GPa) | 30.2 | 36 |
| $E_x$ (GPa) | 65.3 | 72.4 |
| $E_y$ (GPa) | 79.2 | 87.3 |
| $E_z$ (GPa) | 71.1 | 74.3 |
| $\nu_{yz}$ | 0.207 | 0.224 |
| $\nu_{zy}$ | 0.186 | 0.19 |
| $\nu_{xz}$ | 0.269 | 0.337 |
| $\nu_{zx}$ | 0.293 | 0.346 |
| $\nu_{xy}$ | 0.242 | 0.216 |
| $\nu_{yx}$ | 0.293 | 0.261 |
| $A_1$ | 0.695 | 0.79 |
| $A_2$ | 1.2 | 1.38 |
| $A_3$ | 1.055 | 1.11 |
| $A_{Ba}$ | 0.84 | 0.95 |
| $A_{Bc}$ | 0.859 | 1.01 |
| $A_B$ (%) | 0.14 | 0.24 |
| $A_G$ (%) | 2.83 | 3.37 |
| $P_x^{Cauchy} = C_{23}-C_{44}$ (GPa) | -0.66 | -3.11 |
| $P_y^{Cauchy} = C_{13}-C_{55}$ (GPa) | -8.63 | -7.94 |
| $P_z^{Cauchy} = C_{12}-C_{66}$ (GPa) | 7.44 | 9.5 |

Table 5

| Symmetry modes (cm$^{-1}$) | A$_g$ | B$_{1g}$ | B$_{2g}$ | B$_{3g}$ |
|---|---|---|---|---|
| Calculations | 55.26, 76.13, 82.32, 161.96, 165.98, 189.64 | 61.92, 87.27, 120.22, 159.66, 172.82, 177.04 | 48.34, 99.71, 121, 143.92, 169.13, 188.73 | 76.26, 86.06, 113.93, 155.21, 162.45, 183.45 |
| Experiment[61] | 37.5 (very weak), 53, 61, 82, 107, 173 | 53, 66, 150 (shoulder), 178 | 53, 107, 179 | 52, 175 |

Table 6

| Symmetry modes (cm$^{-1}$) | A$_u$ | B$_{1u}$ | B$_{2u}$ | B$_{3u}$ |
|---|---|---|---|---|
| Calculations | 34.67, 45.56, 58.87, 91.99, 152.61, 184.81 | 52.41, 54.64, 118.62, 141.63, 175.87 | 38.79, 57.57, 109.83, 140.4, 183.9 | 58.66, 62.2, 120.04, 153.96, 172.55 |
| Experiment[61] | Silent mode | 58, weak shoulder above 58, 121, 154, 189 | 44, 61, 119, 195 | 66, 123, 166, 184 |

Table 7

| Atom type | $U_{xx}$ (Å$^2$) | $U_{yy}$ (Å$^2$) | $U_{zz}$ (Å$^2$) | $U_{yz}$ (10$^{-4}$Å$^2$) | $U_{zx}$ (10$^{-4}$Å$^2$) | $U_{xy}$ (10$^{-4}$Å$^2$) |
|---|---|---|---|---|---|---|
| Zn | 0.0229 | 0.0184 | 0.02 | -0.34 | 1.12 | 0.29 |
| Sb | 0.0121 | 0.0134 | 0.0141 | -0.35 | 1.14 | 0.29 |

Table 8

| | $v_l$ [m/s] | $v_t$ [m/s] | $v_D$ [m/s] | Θ [K] |
|---|---|---|---|---|
| ZnSb (calculation, this work) at 0 K | 3643.92 | 2095.87 | 2370.65 | 236.1 |
| ZnSb (Experiment) | 3911.67 | 2245 | 2538.17 | 253[40] |
| Zn$_4$Sb$_3$ (Experiment) | 3590 | 2080 | 2310 | 237[2] |
| | 3952 | 2190 | 2470 | 249[74] |

Table 9

| Defect type | $V^{Zn}$ | $V^{Sb}$ | $Sb^{Zn}$ | $Zn^{Sb}$ | $I^{Sb}$ | $I^{Zn}$ |
|---|---|---|---|---|---|---|
| Defect formation energy (eV/def.) | 0.8 | 1.82 | 1.37 | 1.5 | 2.31 | 1.41 |

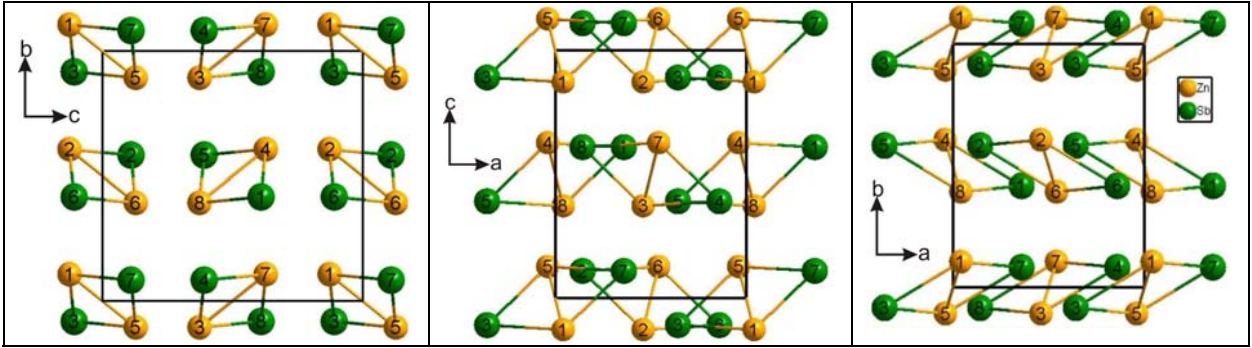

Fig. 1

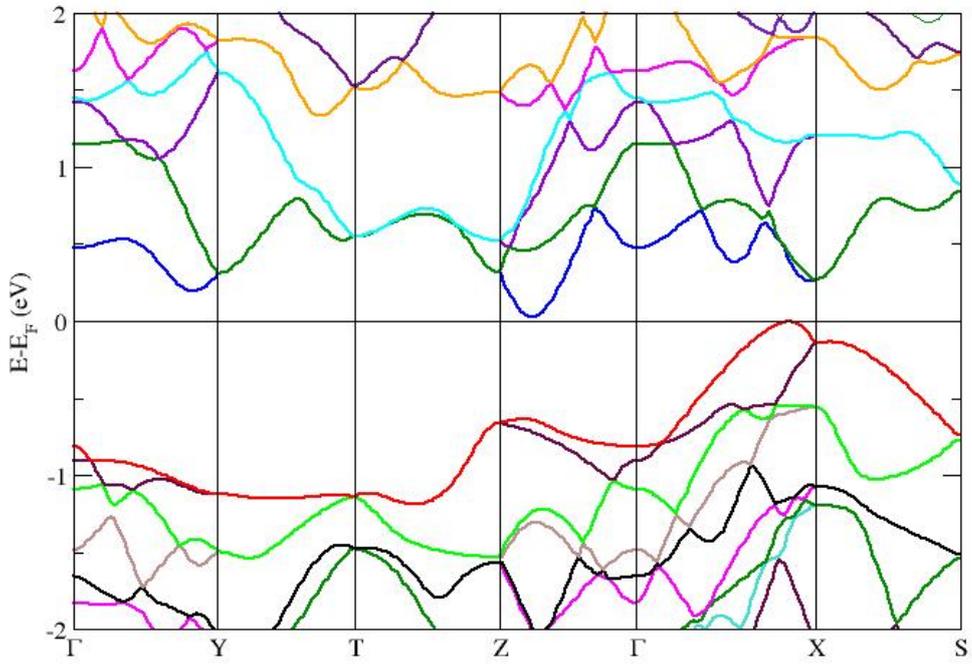

Fig.2

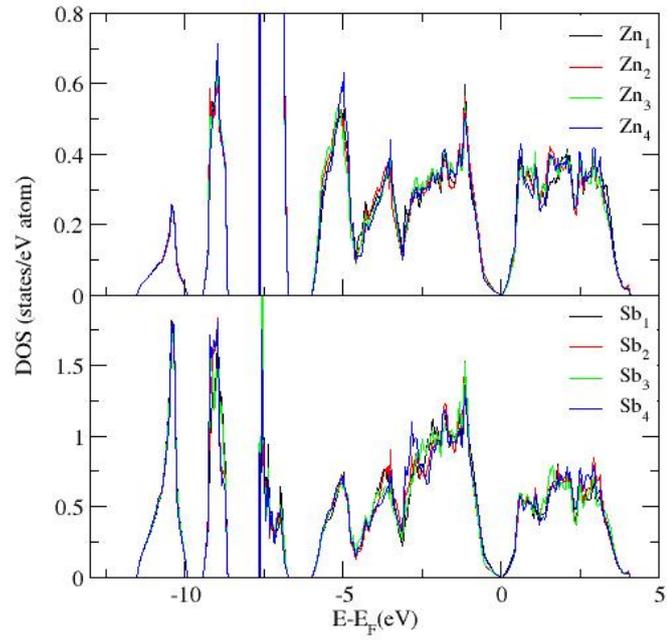

Fig. 3

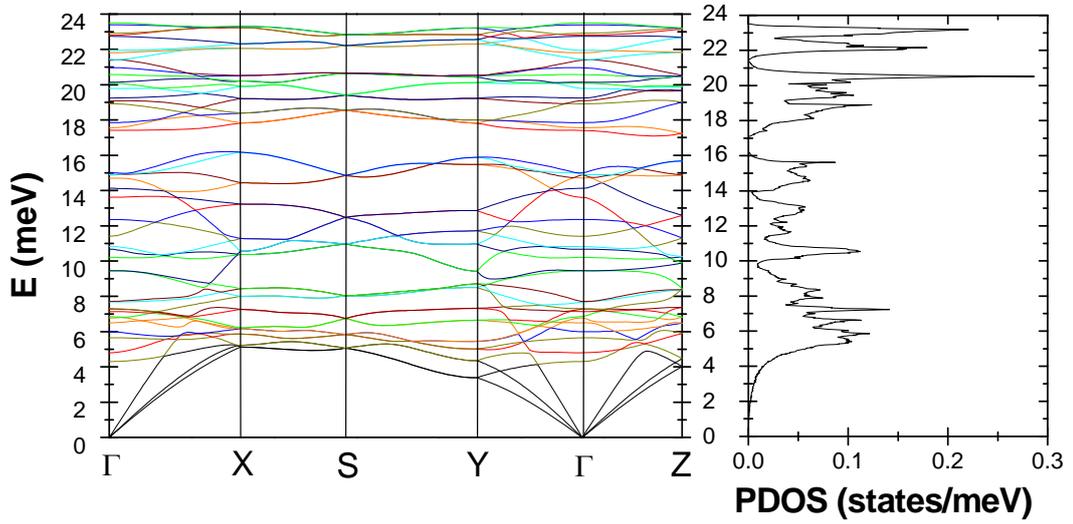

Fig. 4

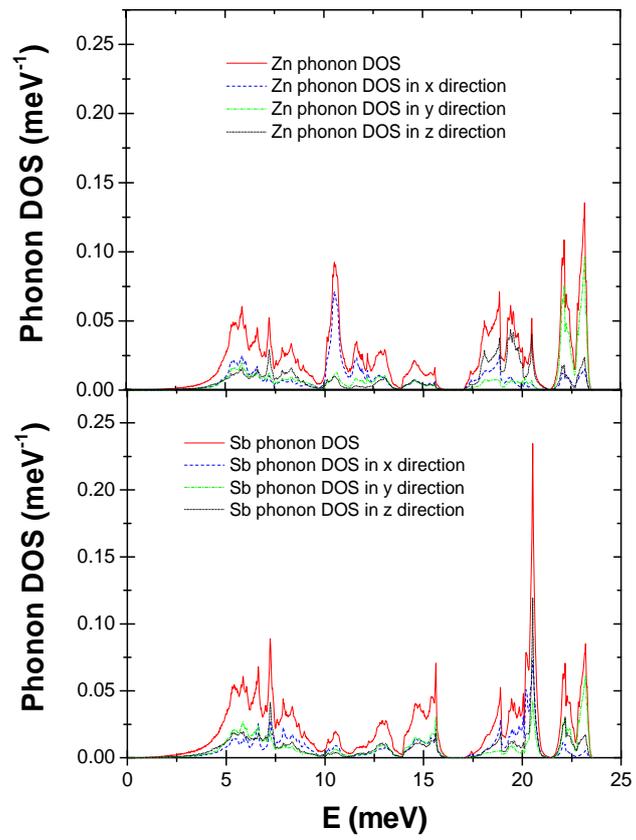

Fig. 5

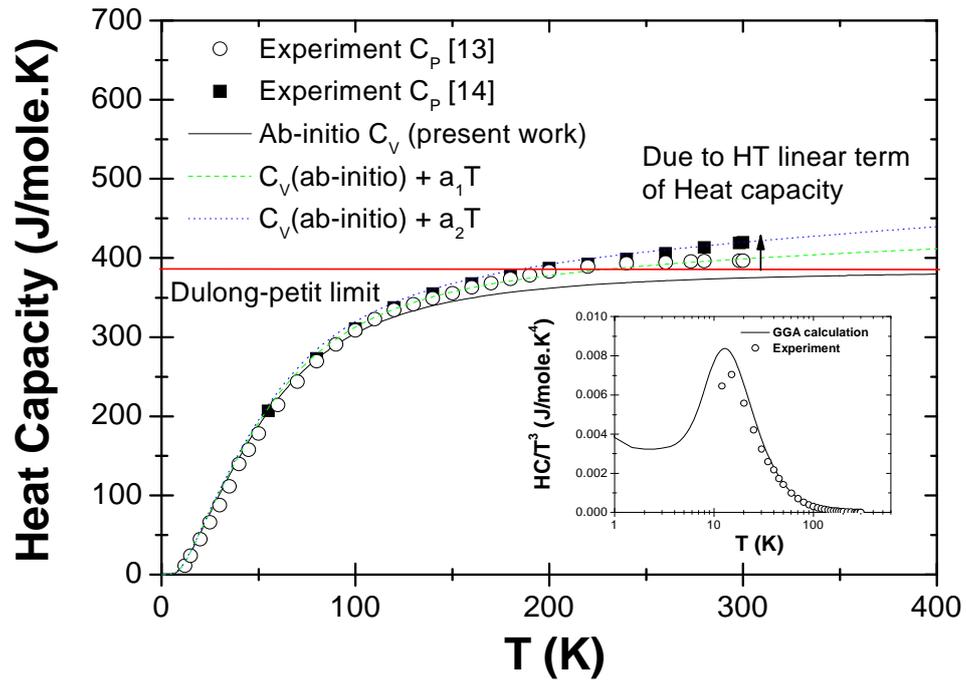

Fig. 6

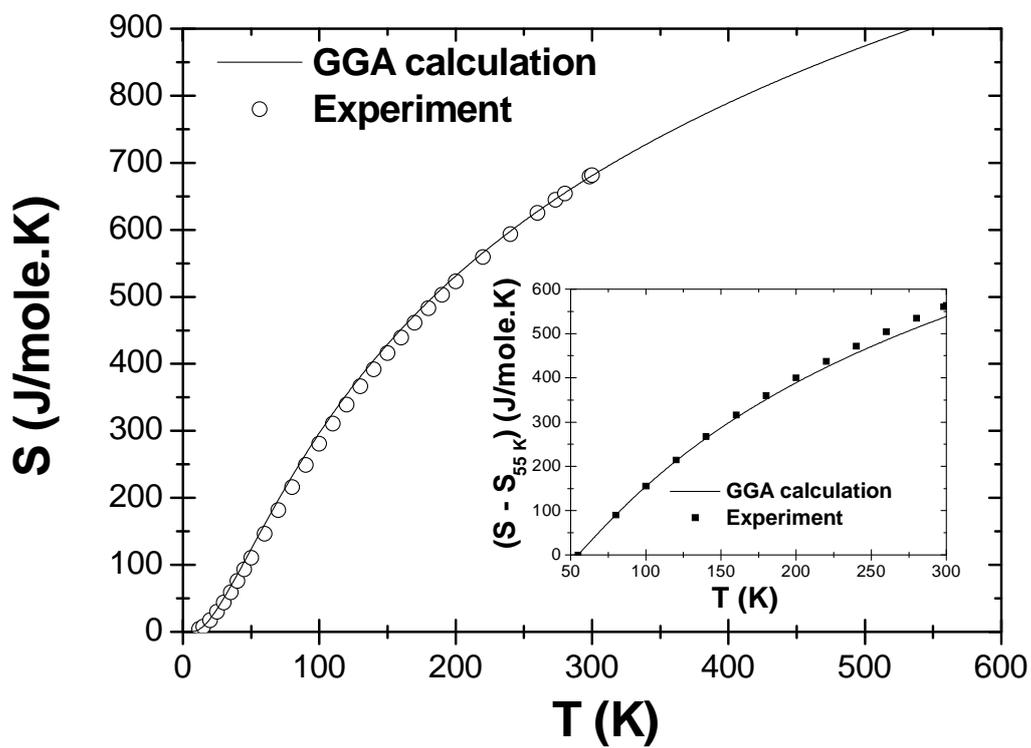

Fig. 7

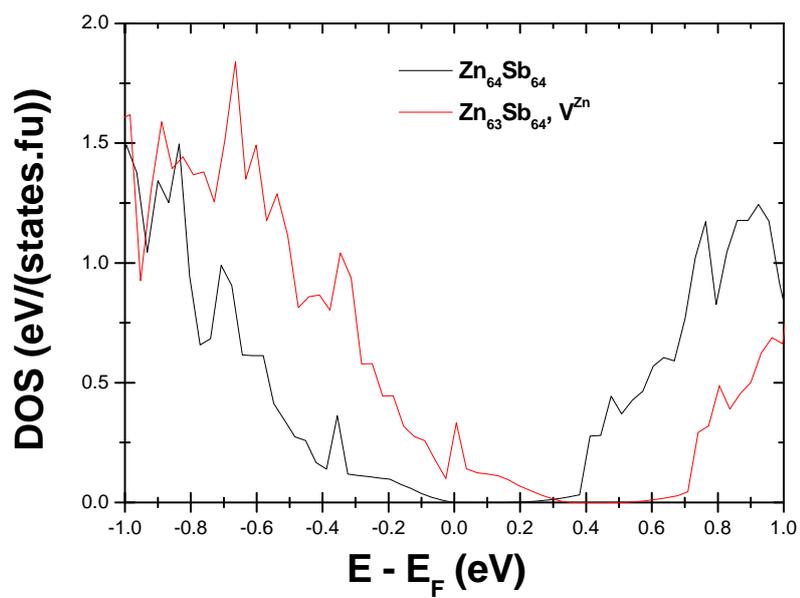

Fig. 8

33